\documentclass{emulateapj}
\usepackage{graphics,graphicx}
\usepackage{psfig}
\usepackage{times}
\usepackage{comment}

\providecommand{\mu}{$\Delta m_{15}(U)$}

\begin{document}
\title{Interpreting Flux from Broadband Photometry}


\author{Peter~J.~Brown\altaffilmark{1}, Alice Breeveld\altaffilmark{2}, Peter W. A. Roming\altaffilmark{3}, \& Michael Siegel\altaffilmark{4} }

\altaffiltext{1}{George P. and Cynthia Woods Mitchell Institute for Fundamental Physics \& Astronomy, 
Texas A. \& M. University, Department of Physics and Astronomy, 
4242 TAMU, College Station, TX 77843, USA; 
pbrown@physics.tamu.edu}            

\altaffiltext{2}{Mullard Space Science Laboratory, University College London, Holmbury St. Mary, Dorking Surrey, RH5 6NT, UK}            
\altaffiltext{3}{Southwest Research Institute, Department of Space Science, 6220 Culebra Road, San Antonio, TX 78238, USA}
\altaffiltext{4}{Department of Astronomy and Astrophysics, The Pennsylvania State University,  525 Davey Laboratory, University Park, PA 16802, USA}

\begin{abstract}

We discuss the transformation of observed photometry into flux for the creation of spectral energy distributions and the computation of bolometric luminosities.  We do this in the context of supernova studies, particularly as observed with the Swift spacecraft, but the concepts and techniques should be applicable to many other types of sources and wavelength regimes. 
Traditional methods of converting observed magnitudes to flux densities 
are not very accurate when applied to UV photometry.  Common methods for extinction and the integration of pseudo-bolometric fluxes can also lead to inaccurate results. The sources of inaccuracy, though, also apply to other wavelengths.  
Because of the complicated nature of translating broad-band photometry into monochromatic flux densities, comparison between observed photometry and a spectroscopic model is best done by forward modeling the spectrum into the count rates or magnitudes of the observations.  
We recommend that integrated flux measurements be made using a spectrum or spectral energy distribution which is consistent with the multi-band photometry rather than converting individual photometric measurements to flux densities, linearly interpolating between the points, and integrating.  We also highlight some specific areas where the UV flux can be mischaracterized.

\end{abstract}


\section{Introduction \label{intro}}

One challenge for astrophysicists (and most scientists in general) is converting observations and theoretical predictions into the same units so that they can be compared.  
Of interest here is the measurement of the intensity of light emitted from astrophysical sources.  The wavelength dependence of the light intensity is usually plotted as flux density versus wavelength.
By flux density, we mean the energy of light  from a unit of wavelength given as erg cm$^{-2}$ s$^{-1}$ \AA$^{-1}$ (or frequency in units of erg cm$^{-2}$ s$^{-1}$ Hz$^{-1}$).  
Flux is the integral of the flux density over a region of wavelength or frequency.  
Measuring such a fundamental parameter as flux or flux density is complicated because modern detectors are usually sensitive to the number of incoming photons, rather than the amount of incident energy flux. Forward modeling from theory to observations is preferred when possible.  The inverse problem is much more difficult because a myriad of flux spectra could reproduce the limited quantities constrained by photometric observations.  Conversions of a photometric magnitude back into a physical flux, for example, is non-trivial when the broadband filter covers a range of different energies and the source spectrum is unknown.
To understand the energetics involved and to compare with theoretical models, 
it is often desirable to measure what is called the bolometric flux (or luminosity) -- namely, the total energy flux received (or luminosity emitted) by an object across all energies. 
Bolometric luminosity is an important observational property because they can be compared to theoretical models without requiring accurate radiative transfer models to predict the output spectrum \citep{Vacca_Leibundgut_1996}.

The true bolometric flux is impossible to measure directly.  Bolometric flux or magnitudes can be estimated utilizing observed magnitudes in one or more bands and ``bolometric corrections'' based on stellar models or blackbody spectra \citep{Stromberg_1932,Bleksley_1935}.
The earliest estimate of the bolometric flux of a supernova (SN) was based on a blackbody curve fit to the optical luminosity of SN~1885 in Andromeda \citep{Baade_Zwicky_1934}.  
A ``pseudo-bolometric'' flux measurement tries to capture a significant fraction of the light and can be computed in many different ways.  
Sometimes the flux is integrated directly from spectrophotometry \citep{Code_etal_1976}.
A common method involves transforming observed magnitudes into monochromatic flux densities and ``connecting the dots'' with linear segments or a spline fit (e.g. \citealp{Suntzeff_Bouchet_1990,Stanishev_etal_2007}).  
Another method calculates the flux from each filter by multiplying the mean flux by the effective width of the passband \citep{Vacca_Leibundgut_1996}.  Gaps and overlap between filters are accounted for in adding up the total flux.
Sometimes the flux outside of the observed bands is accounted for as a percentage of the observed flux \citep{Vacca_Leibundgut_1996}.  Other details and methods will be discussed further below.

Observing the largest possible wavelength range allows the bolometric flux to be more accurately determined by reducing the uncertainty on the unobserved flux.  However, as observations stretch to much higher and shorter energies, the same techniques and methods may no longer be appropriate.
The launch of the Swift satellite with its Ultra-violet/Optical Telescope (UVOT; \citealp{Gehrels_etal_2004,Roming_etal_2005}) has led to an explosion in time-series UV data on SNe which can be incorporated into bolometric light curves.  It is appropriate to reassess the assumptions and techniques used in calculating bolometric luminosities and evaluate their appropriateness.
In this paper we focus on the region covered by Swift UVOT (1600-6000 \AA), but the principles should be more widely applicable. 
Some of the critique given is in reference to current computation of bolometric light curves of SNe, whose spectral flux changes rapidly across the wavelength range of the UV filters.  Similar complications may arise for other source-filter combinations, such as optical observations of very cool stars or photometric observations of stars with large molecular bands, steep spectral slopes, or other large features within the observed bandpass. 
 Section  \ref{conversion} covers the more general issue of the conversion of observed magnitudes or count rates to a flux density spectral energy distribution (SED).   The correction for extinction is addressed in Section \ref{extinction}.  In Section \ref{bolo} we will discuss the wavelength limits and integration methods used for a bolometric or integrated luminosity measurement and Section \ref{comparisons} compares the results using different methods. We present integrated flux measurements for a sample of SNe in Section \ref{snbolo}.  In Section \ref{conclusion} we summarize and give our recommendations.

 
\section{Converting Observations to Flux Densities} \label{conversion}

For many comparisons with observed or theoretical spectra, it is straightforward to integrate the product of a spectrum and the wavelength-dependent system transmission (including filter, detector and atmospheric effects) to get a value which can be compared to observed photometry.  The consistency of different models can be compared using the $\chi^2$ values or other statistical tests.  But often a visual representation is desired in addition to the statistics, so one wants to plot the photometry on the spectrum or create a wavelength-flux spectral energy distribution (SED)\footnote{In this paper the term SED will refer to a low resolution spectrum such as that constructed from multiple broad-band photometric measurements.}.  The measured count rate or magnitude through a particular filter needs to be transformed into a wavelength and a flux density.  
 As stated in \citet{Davis_Webb_1970},  ``The use of monochromatic fluxes at the effective wavelengths of the observations for the comparison, rather than fluxes obtained by folding the various sensitivity functions through the models, is justified by the linearity of the model continua over the experimental passbands.''  
The use of monochromatic flux densities is now quite common regardless of the continua shape, and the limits and errors of such methods are not usually addressed.  Before discussing the advantages and disadvantages of specific techniques, we wish to first emphasize that a broadband measurement is affected by the original source spectrum, reddening from intervening dust (local to the source, intergalactic at a range of relative velocities, and Milky Way), the Earth's atmosphere (for ground-based observations), the instrumental efficiency (including mirror and lens reflectivity or transmission), filter throughput, and detector sensitivity.  Many of these have a wavelength dependence.  Determining the original flux which resulted in the observed count rates requires assumptions or corrections for these effects in either the photometric calibration or flux conversion.  We wish to draw attention to many of these assumptions and corrections and encourage others to assess the importance of each for their particular circumstance.

There are many possibilities for the reference wavelength to use for a filter: the wavelength of peak transmission, or the central, mean, isophotal or other characteristic wavelengths used to define a filter when astronomers studied in detail how to interpret broad-band measurements for different spectral shapes (see e.g. \citealp{Golay_1974}).  
Here we will use the spectral weighted effective wavelength defined below, where E$\lambda$ and S$\lambda$ are the filter transmission and spectral flux density as a function of the wavelength:
\begin{equation}
\lambda = \int [ \lambda E(\lambda) S(\lambda) d\lambda ]  / \int [ E(\lambda) S(\lambda) d\lambda ]
\end{equation}
These effective wavelengths are not just a function of the filter transmission but also the source spectrum, so the continuum shape, strong absorption features \citep{Siegel_etal_2012}, and reddening \citep{Brown_etal_2014J} can affect it.  It can be a useful diagnostic of the wavelengths from which the detected photons are coming.

The conversion of observed magnitudes (or the actual observed photon or electron count rates) to a flux density is one of the most fundamental calculations.  Yet the methods described vary and are sometimes considered too trivial to describe.  Several conversion factors have been published over the decades that are applicable to  ``standard'' systems  \citep{Johnson_1966,Bessell_1979,McWilliam_1991} or for a specific instrument system (e.g. \citealp{Poole_etal_2008}).  The actual conversions used when plotting SEDs, however, are not always cited nor is the applicability of those conversions frequently discussed, despite their dependence on the filter/detector characteristics and the spectral shape of the source. 
The SN community has realized the need to take actual filters into account when comparing photometry from different systems rather than just using color terms based on standard stars very different than SNe \citep{Suntzeff_2000}.  This is formalized in the use of filter and spectrum-dependent `s corrections' \citep{Stritzinger_etal_2002}.  The same corrections with regards to flux conversion have yet to be widely recognized.

\begin{figure} 
\resizebox{7.0cm}{!}{\includegraphics*{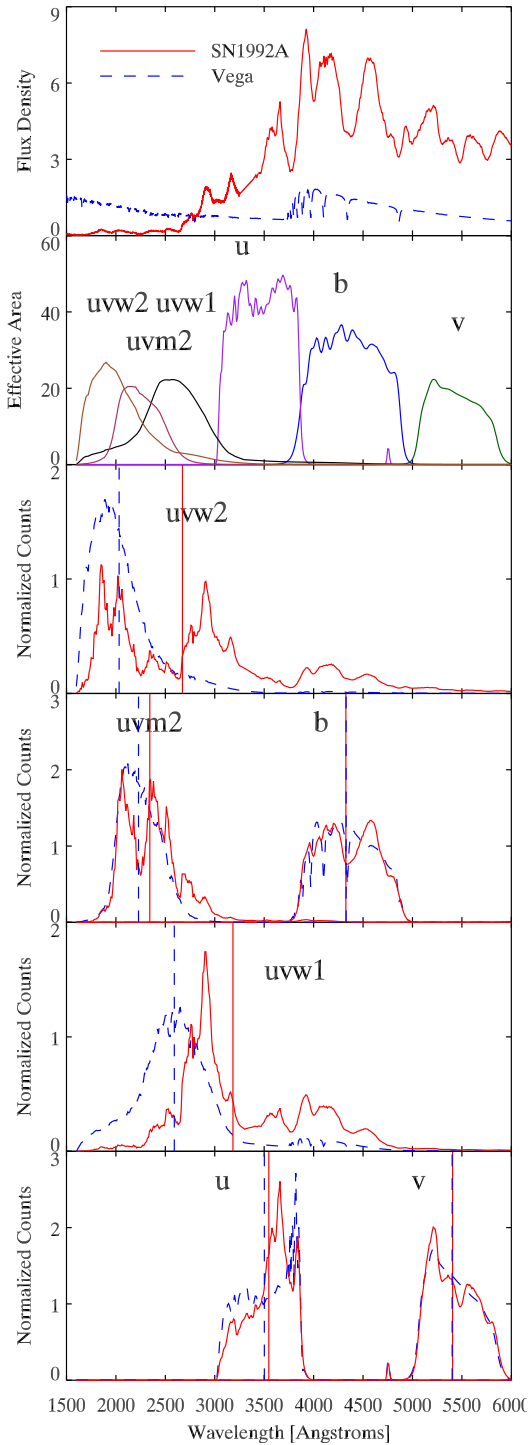}   }
\caption[Results]
        { Top Panel: Spectra of Vega and the type Ia SN~1992A.  Second Panel: Effective area curves of the Swift/UVOT filters.  Lower panels: The transmitted counts through the Swift/UVOT filters (multiplying the flux density by the effective area and converting from flux to counts) for Vega and SN~1992A.  The effective wavelengths for SN~1992A and Vega  are plotted with solid and dotted vertical lines, respectively.  The largest differences are for the uvw2 and uvw1 filters.
 } \label{fig_countspectra}    
\end{figure}

The true relationship between observed count rates and flux density is complicated by differences in spectral shapes and the finite width of filter bandpasses.  As stated by \citet{Golay_1974}, "So we see that heterochromatic photometry (i.e., using a non-negligible passband) can theoretically only provide information about the function E($\lambda$) at a point $\lambda_0$ when the energy distribution contains no lines and when the slope of the continuum does not vary too rapidly with $\lambda$."  For most photon-based systems, where an incoming photon is converted into an electrical signal of some sort, the detector does not know the energy of that photon.  A photon at the highest energy allowed through the filter is counted the same as a photon at the lowest energy transmitted.  Figure \ref{fig_countspectra} contrasts the flux spectra, and resulting count spectra, of Vega\footnote{The spectrum used is archived at ftp://ftp.stsci.edu/cdbs/calspec/alpha\_lyr\_stis\_004.fits} \citep{Bohlin_Gilliland_2004} and the type Ia SN 1992A \citep{Kirshner_etal_1993} if observed through the Swift/UVOT filters.  Also shown are the effective wavelengths for each filter and spectrum combination.  While the count spectra are not too dissimilar in the optical, the count distributions and effective wavelengths diverge in the UV.  The expected flux at the effective wavelength varies with spectral shape.

\begin{figure*} 
\resizebox{18cm}{!}{\includegraphics*{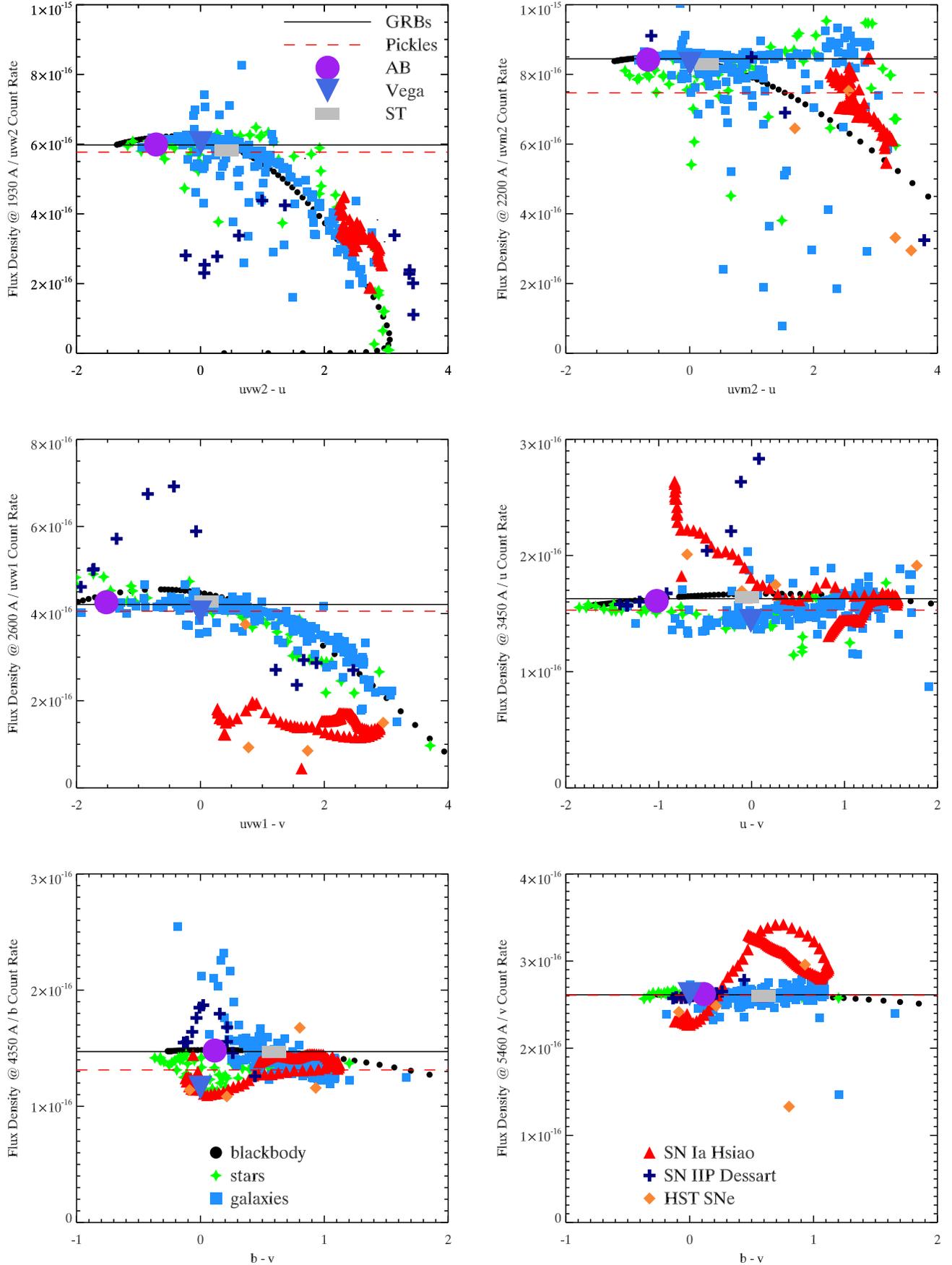}}
\caption[Results]
        {Conversion factors between the observed count rates and the flux density at the Vega effective wavelengths for the six Swift/UVOT filters.
 } \label{fig_fluxfactors}    
\end{figure*}

A source-specific determination of those factors, however, can reduce the uncertainty on the flux conversion and reduce systematic errors in the derived flux.  \citet{Brown_etal_2010} show the variation in those conversion factors for the UVOT filters over a wide range of stellar spectral types, SN templates, galaxy models, and blackbody spectra. 
In Figure \ref{fig_fluxfactors} we show similar plots for each of the six broad-band filters used for SN observations\footnote{Plots showing the ``white'' count rate to flux conversion factors versus source color are shown in the UVOT calibration database (CALDB) documentation at http://heasarc.gsfc.nasa.gov/docs/heasarc/caldb/swift/docs/uvot/uvot\_caldb\_counttofluxratio\_10wa.pdf}.  Stellar spectral models come from \citet{Pickles_1998}, galaxy spectra from \citet{McCall_2004} and \citet{Brown_etal_2013}, an average Type Ia SN spectral series from \citet{Hsiao_etal_2007}, and a Type IIP SN synthetic spectral series from \citep{Dessart_etal_2008}.  Also plotted are the flux conversions for GRB, stellar spectra, and AB models from the Swift/UVOT CALDB documentation (first determined in \citealp{Poole_etal_2008} but updated for the revised UV filter curves of \citealp{Breeveld_etal_2011}).  We also plot the conversion factors for a spectrum flat in flux density per unit wavelength (like in the STMAG system\footnote{The UVOT zeropoints in the STMAG system for the uvw2, uvm2, uvw1, u, b, and v filters are 16.99, 16.60, 17.32, 18.36, 18.49, and 17.86, respectively.)} as in \citealp{Koornneef_etal_1986}).  The flux conversions factors are tabulated in Table 1. 

The variation in the optical is small for many source types, but can vary by a factor of several. 
The UV flux conversion factors vary by over an order of magnitude.  This is due in part to dramatic changes in the spectral shapes over the range of the filters and the large difference in energy between photons transmitted through the ends of the filters.  For the uvw2 and uvw1 filters this is exacerbated by long optical tails in the throughput curves which transmit flux over a large wavelength range.  The strong effects in uvm2 are, however, not caused by significant red leaks.  One should understand these differences (whether considered as a change in the conversion factors or a shift in the effective wavelengths) rather than just dismissing it as a red leak issue.  Converting count rates from the Swift/UVOT ``white'' filter is even more complicated due to its very wide passband.  Other space or ground-based filters might have similar issues due to particular spectral shapes or the broadness of the filters.  

\begin{figure*} 
\resizebox{18cm}{!}{\includegraphics*{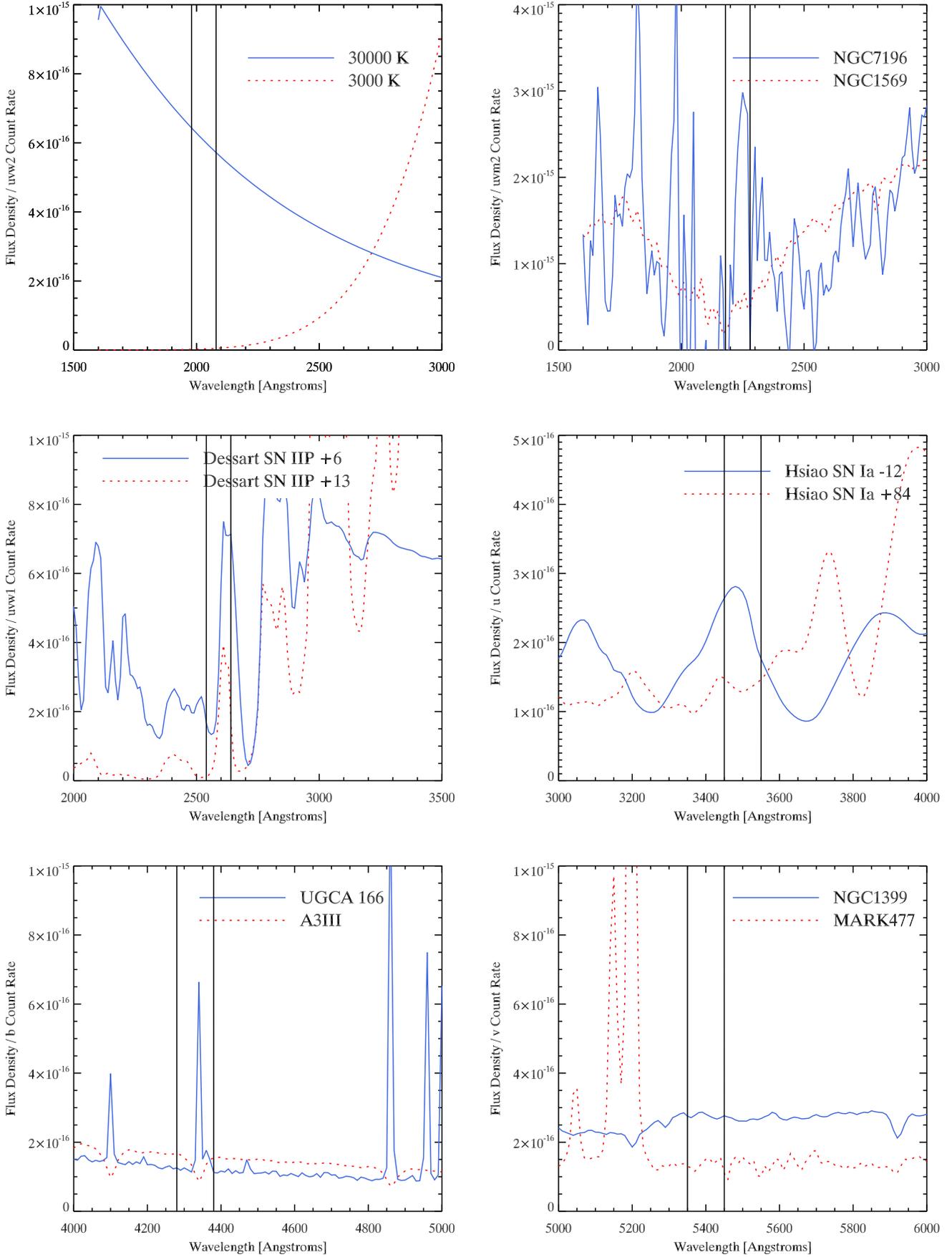}}
\caption[Results]
        {Flux density spectra are divided by the count rates in the six UVOT filters to show the variety of spectral shapes which could have the same magnitude yet wildly different flux densities at the Vega effective wavelengths.  Vertical lines denote the bounds within which the flux density is computed for the conversion factors.
 } \label{fig_fluxfactorspectra}    
\end{figure*}

Figure \ref{fig_fluxfactorspectra} illustrates some of the causes of these flux conversion differences.  In each panel, a flux density is normalized by the observed count rate through one of the six UVOT filters.  Thus in each panel, both spectra would have the same observed magnitude.  Nevertheless, the shape of the spectra and the value of the flux density at the Vega effective wavelength can be quite different.  The flux densities vary because of very different spectral shapes and also strong absorption or emission features close to or outside the effective wavelength range used for computing the conversion factors.  This highlights the need to be careful in interpreting broadband photometry as a monochromatic flux density even in the optical.  While we give flux conversion factors for a variety of sources in Table 1, we caution that their usefulness is limited.  Below we discuss a few specific uses of broadband flux densities and discuss better conversion methods.

\begin{deluxetable}{llrrrrrr}

\tablecaption{Flux Conversion Factors}

\tablenum{1}

\tablehead{\colhead{Spectrum} & \colhead{References} & \colhead{uvw2} & \colhead{uvm2} & \colhead{uvw1} & \colhead{u} & \colhead{b} & \colhead{v} \\ 
\colhead{} & \colhead{} & \colhead{} &  \colhead{} & \colhead{} & \colhead{} & \colhead{} & \colhead{} } 

\startdata
Vega & (1) & 6.03 & 8.30 & 4.02 & 1.44 & 1.16 & 2.62 \\
GRBs & (2)  & 5.98 & 8.45 & 4.21 & 1.63 & 1.47 & 2.61 \\
Pickles & (3)  & 5.77 & 7.47 & 4.06 & 1.53 & 1.31 & 2.61\\
AB  & (4)  & 6.23 & 8.49 & 4.63 & 1.66 & 1.48 & 2.61 \\
ST  & (5)  & 6.03 & 8.30  &  4.02  & 1.44  & 1.16 & 2.62 \\
3000 K & (6) &  0.03 &  2.53 &  0.57 &  1.49 &  1.31 &  2.53 \\
10000 K & (6) &  5.76 &  8.28 &  4.55 &  1.64 &  1.49 &  2.62 \\
30000 K & (6) &  6.06 &  8.42 &  4.02 &  1.57 &  1.48 &  2.62 \\
a0i & (7) &  6.24 &  7.53 &  4.01 &  1.37 &  1.34 &  2.63 \\
a0iii & (7) &  5.86 &  7.56 &  4.42 &  1.45 &  1.14 &  2.62 \\
a0v & (7) &  6.05 &  7.94 &  4.15 &  1.44 &  1.19 &  2.62 \\
g0v & (7) &  0.09 &  6.72 &  2.89 &  1.61 &  1.33 &  2.60 \\
o9v & (7) &  6.08 &  7.96 &  4.29 &  1.58 &  1.40 &  2.58 \\
g1050 04 & (8) & 6.97 &  8.87 &  3.55 &  1.37 &  1.67 &  2.51 \\
ic3639 & (8) & 6.09 &  6.97 &  3.82 &  1.47 &  1.52 &  2.53 \\
mrk477 & (8) & 6.83 &  8.98 &  4.32 &  1.32 &  1.22 &  1.46 \\
ngc6221 & (8) & 1.61 &  5.23 &  3.83 &  1.59 &  1.40 &  2.69 \\
ngc7496 & (8) & 7.42 &  7.90 &  4.17 &  1.49 &  1.40 &  2.66 \\
IC 4051 & (9) & 2.50 &  8.93 &  2.94 &  1.65 &  1.30 &  2.57 \\
IC 5298 & (9) & 5.75 &  8.07 &  3.99 &  1.53 &  1.36 &  2.57 \\
II Zw 096 & (9) & 6.27 &  8.54 &  4.24 &  1.45 &  1.58 &  2.64 \\
NGC 0520 & (9) & 5.52 &  8.44 &  3.84 &  1.61 &  1.31 &  2.62 \\
NGC 0584 & (9) & 2.30 &  8.92 &  2.96 &  1.56 &  1.37 &  2.63 \\
Hsiao 0 & (10) & 3.67 &  6.81 &  1.91 &  2.15 &  1.14 &  2.28 \\
Hsiao 15 & (10) & 2.52 &  6.11 &  1.24 &  1.65 &  1.28 &  3.13 \\
SN05cs+3 & (11) & 2.81 &  9.11 &  4.61 &  1.58 &  1.55 &  2.57 \\
SN05cs+17 & (11) & 1.11 &  1.40 &  2.70 &  3.13 &  1.26 &  2.78 \\
Ia SN2011fe & (12) &  3.20 &  2.95 &  0.93 &  2.01 &  1.14 &  2.42 \\
Ia SN1992A & (13) &  3.82 &  3.32 &  0.85 &  1.75 &  1.08 &  2.48 \\
Ic SN1994I & (14) &  3.40 &  7.53 &  1.50 &  1.91 &  1.16 &  2.96 \\
IIP SN1999em & (15) &  5.59 &  6.45 &  3.75 &  1.69 &  1.68 &  1.33 \\
\enddata

\tablecomments{Conversion factors are multiplied by the count rate to give the flux density in units of 10$^{-16}$ erg s$^{-1}$ cm$^{-2}$ \AA$^{-1}$.  Table 1 is published in its entirety in the machine-readable format.
      A portion is shown here for guidance regarding its form and content.}

\tablerefs{
(1) Vega spectra from ftp://ftp.stsci.edu/cdbs/calspec/alpha\_lyr\_stis\_004.fits \citep{Bohlin_Gilliland_2004}.
(2) The value given is the average computed for a variety of GRB models described in \citet{Poole_etal_2008}.
This value is used in the Swift CALDB products.
(3) The value given is the average computed for a variety of stellar spectra \citep{Pickles_1998} as described in \citet{Poole_etal_2008}.
(4) This value is given in the Swift/UVOT CALDB documentation for the  AB magnitude system as defined by \citet{Oke_1974}. 
(5) The STMAG system is based on a spectrum with constant flux density per unit wavelength as described by \citet{Koornneef_etal_1986}.
(6) Blackbody spectrum calculated according to \citet{Planck}.
(7) Stellar spectra from \citep{Pickles_1998}.
(8) Galaxy spectra from \citet{Storchi_etal_1995}
(9) Galaxy spectra from \citet{Brown_etal_2013}
(10) Average type Ia SN spectral series from \citet{Hsiao_etal_2007}.  The number indicates days from maximum light (positive or negative).
(11)  Theoretical spectra matched to Type IIP SN~2005cs from \citet{Dessart_etal_2008}.  The number indicates days from explosion.
(12) Type Ia SN2011fe spectra from \citet{Mazzali_etal_2014}.
(13) Type Ia SN1992A spectrum from \citet{Kirshner_etal_1993}.
(14) Type Ic SN1994I spectrum from \citet{Jeffery_etal_1994}.
(15) Type IIP SN1999em spectrum from \citet{Baron_etal_2000}.
}
\end{deluxetable}

\section{Comparisons of Photometry with Spectral Models}

\begin{figure} 
\resizebox{7.6cm}{!}{\includegraphics*{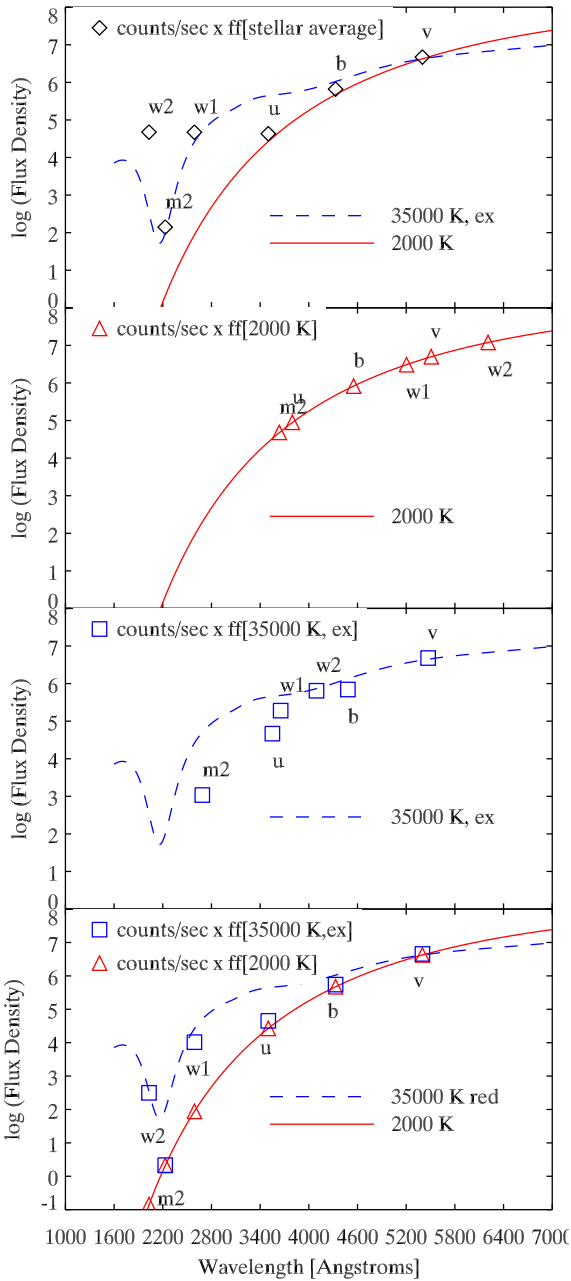}}
\caption[Results]
        {SED reconstructions from spectrophotometry (in the Swift/UVOT system) are compared to models.
Top Panel: The spectra of a 2000 K blackbody and a reddened 35000 K blackbody are compared to an SED made by converting the 2000 K blackbody spectrophotometry to flux using the standard flux conversion factors (abbreviated as ``ff'') and effective wavelengths.  
Second Panel: The 2000 K blackbody spectrophotometry is converted to flux using the blackbody spectrum to derive the conversion factors and the effective wavelengths expected from that model.   
Third Panel: The 2000 K blackbody spectrophotometry is converted to flux using the 35000 K reddened blackbody spectrum.  The photometry and model are clearly inconsistent. 
Bottom Panel:  An alternative visualization of the photometry, in which the plotted wavelengths (at the Vega effective wavelengths of the Swift/UVOT filters) are held constant and the appropriate conversion factors calculated for different spectral models and compared to the spectral models themselves.  Flux is not actually measured, but is computed in a model-dependent way that must be done correctly.
 } \label{fig_flux}    
\end{figure}

An appropriate flux conversion can be computed for a given spectrum and may be useful for visualizing the spectral shape of an object or comparing flux ratios at given wavelengths. 
But testing the validity of a model by comparing the converted flux density to the flux density of a model spectrum adds unnecessary conversions and assumptions.    Forward modeling is a more straightforward way to compare models and photometry with the assumptions made plain.  Synthetic photometry should be performed on the spectrum (including any extinction or other spectral-dependent corrections) and directly compared to the observed photometry.  If a spectral visualization is desired, one can use the model spectrum being compared against to compute the spectrum-dependent effective wavelength (if desired) and the flux conversion for the wavelength being used.

Figure \ref{fig_flux} illustrates the effect that incorrect flux conversions can have on the selection of a best-fit model with the following example.  We begin with a 2000 K blackbody spectrum and compute spectrophotometry in the UVOT system.  The photometry is converted to flux densities using the average conversions from \citet{Poole_etal_2008}.  
The top panel of Figure \ref{fig_flux} shows the input spectrum along with the computed flux.   The computed flux densities match the flux of the spectrum in the optical but not the UV, with the uvw2 and uvw1 fluxes in particular being much higher.  We find the UV flux to be a reasonable match to a hotter blackbody (35000 K) with high reddening (E(B-V)=2.3 with a Milky Way extinction law with R$_V$=3.1 using the \citet{Cardelli_etal_1989} parameterization). 
The flux in the uvw2 and uvw1 filters is dominated by the optical light for very red sources, so the standard correction factors overestimate the UV flux.  This is because the conversion implicitly assumes the same fraction of UV to optical photon counts as in the spectra used to compute the conversion factors.  The uvm2 filters is less affected, resulting in a dip reminiscent of Milky Way extinction.

If the actual spectrum is known (or a suitably accurate template is found via $\chi^2$ comparison with the photometry), the effective wavelengths and flux conversions can be exactly determined.   For a red spectrum the effective wavelengths shift strongly to longer wavelengths.  The comparison is shown in the second panel of Figure \ref{fig_flux}--an exact match by construction.   In the third panel, the conversions and comparisons are made with the hotter, reddened blackbody curve.  The fluxes derived from the photometric points are clearly discrepant.  The information (i.e. consistency in panel 2 but not panel 3) is not new-- a direct comparison of the observed photometry and the model spectrophotometry can already distinguish between the models--but is a visually reassuring way to compare the observations and the models.

An alternate visualization is shown in the bottom panel, where the effective wavelengths are fixed to the Vega effective wavelengths and only the conversion factors recalculated from the respective spectra.  For the UV filters with red tails, a red spectrum is accounted for not by a shift in the plotted wavelength but an appropriately small flux conversion factor to account for most of the counts coming from optical photons.  The triangle symbols are consistent with the solid-line cool spectrum, while the squares are not consistent the dashed-line spectrum.  

It may seem circular to assume a spectrum shape to convert the flux and determine if the photometry is consistent with the spectrum.  However, most comparisons assume an average stellar spectrum, a Vega spectrum, or a flat spectrum (AB) to compute the flux, none of which is likely correct.  Thus even if the flux agrees, the accuracy is still in question because it was assumed that the spectra were different.  If the spectrum under question is assumed, they can at least be shown whether they are consistent, while a disagreement means that the spectrum is wrong rather than just being an error in the assumptions.  Even if they are consistent in total counts (magnitudes) the flux could be different because not all counts have the same energy, but this is certainly better than having an SED not consistent with the observations and assuming the flux somehow comes out correct.  Thus this method can be used to falsify a model but not conclusively validate it.  A more straightforward comparison is to just compute a synthetic magnitude from the model spectrum to compare with the observed photometry.  This would naturally fold in the filter characteristics (including any optical tails).

\section{Extinction Correction} \label{extinction}

An important component of calculating a bolometric luminosity is correcting for the line of sight extinction, whether from dust in the Milky Way, interstellar dust, or in the host galaxy of the SN (each of which could have a different wavelength dependence).
Correcting for extinction in the UV is an extremely complicated subject.  The wavelength dependence of extinction varies with location in the Milky Way.  Small differences in the assumed reddening, e.g. the difference between the Milky Way reddening inferred by \citet{Schlegel_etal_1998}, \citet{Schlafly_Finkbeiner_2011}, or \citet{Planck_etal_2015_dust}, 
can result in apparent differences \citep{Peek_Schiminovich_2013}.  It could also vary between galaxies and in the circumgalactic medium \citep{Peek_2013}. We do not make any claims on the correct extinction law to use.  We do show that even if the extinction law is precisely known, an inappropriate application of that law can have significant consequences.

Typically, broad-band photometry is converted to a flux and corrected for extinction using the R$_\lambda$ from the \citet{Cardelli_etal_1988} or other such extinction law \citep{Pei_1992, Fitzpatrick_Massa_2007, Gordon_etal_2003} computed at the effective wavelength of the associated filter.  This may be generally adequate in the optical, where the extinction laws are smooth and monotonic, but may not be accurate if the spectrum is strongly varying or there are strong emission lines \citep{Clocchiatti_etal_2008}.  In the UV, the strongly varying shapes of both the reddening functions and the source spectrum continuum mean that the effective reddening coefficient R for a given filter depends strongly on the source spectrum and the total amount of reddening \citep{Brown_etal_2010,Brown_etal_2014J}.  In particular, the Swift uvm2 filter sits right on top of the 2175 \AA~bump in the Milky Way extinction curve \citep{Cardelli_etal_1989}.  This higher than average extinction is then assumed to apply to the whole bandpass most of which has a correction factor below that.    An extreme case would be a type Ia SN like SN~1992A.  It features a flux deficit at the same location as the extinction curve (believed to be intrinsic rather than caused by an extinction bump).  Because there is so little flux where the extinction is strongest, the effective extinction is lower because the extinction is lower at the wavelengths where there is actually flux.  Because extinction laws generally redden spectra, bluer spectra are more efficiently reddened such that more flux is lost for the same amount of optical extinction or color excess.  

When calculating a bolometric luminosity, the problems of flux conversion discussed above, such as the overestimate of UV flux for very red spectra, are also exacerbated for situations of high reddening.   For intrinsically red spectra, this results in a negligible overestimate of the integrated flux.  In the case of a reddened spectrum, however, the UV flux is first overestimated and then multiplied by a large correction factor.  

\begin{figure*} 
\plottwo{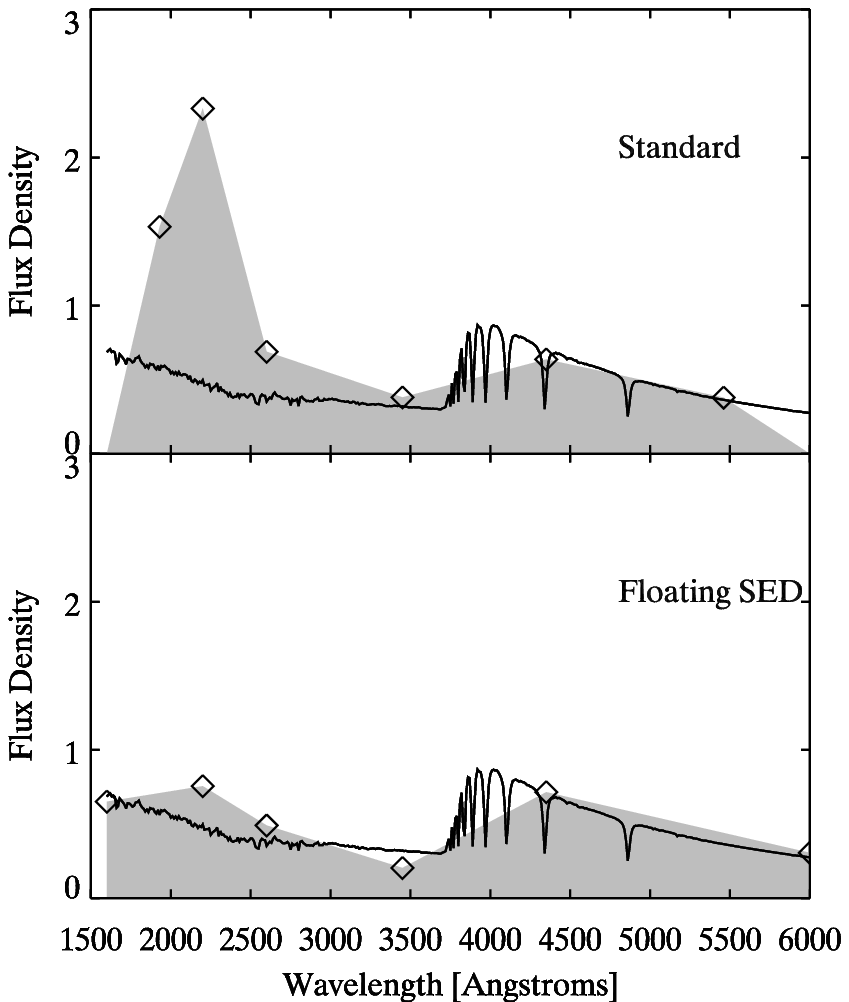}{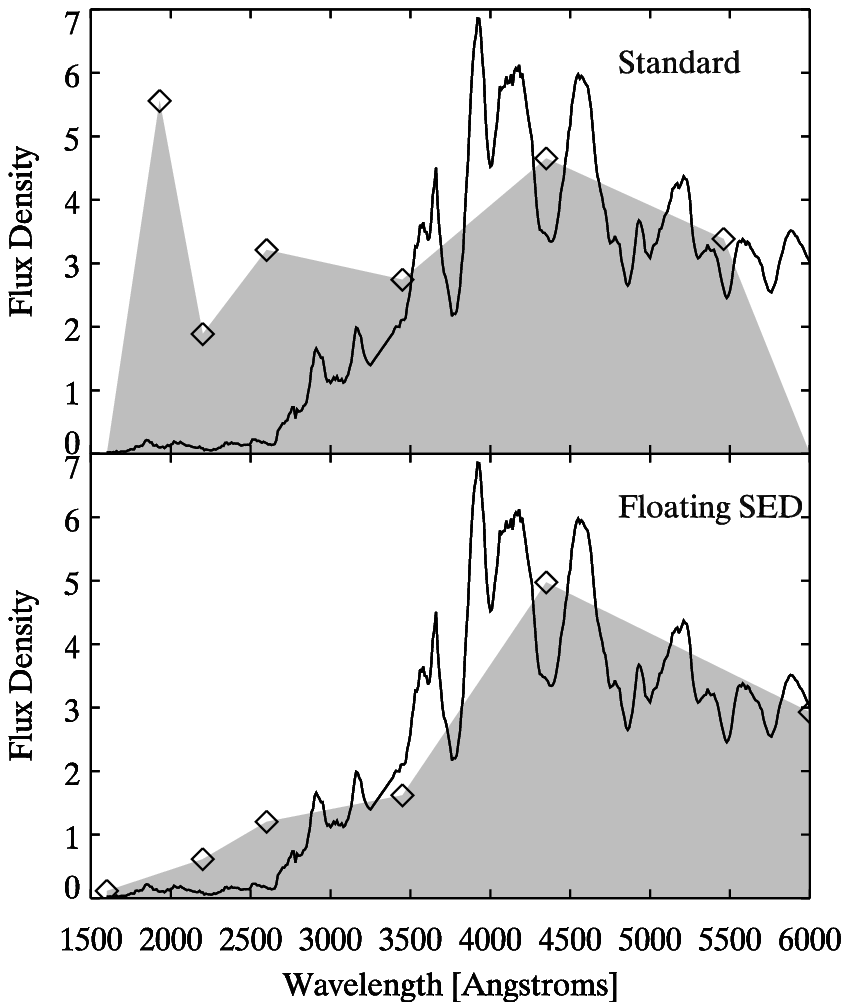}  
\caption[Results]
        {SED reconstructions (in units of ergs s$^{-1}$ cm$^{-2}$ \AA$^{-1}$) are made from reddened photometry of Vega (left panel) and SN~1992A (right panel) and then corrected for reddening for comparison with the original spectrum.
In the top panels, the standard flux conversion is performed and then corrected for extinction.  In the bottom panels, an SED is created which is photometrically consistent with the reddened photometry and then corrected for extinction.  
 } \label{fig_redseds}    
\end{figure*}

In Figure \ref{fig_redseds} we show this effect with Vega and SN~1992A.  First we redden the  input spectrum with a Cardelli Milky Way extinction law with R$_V$=3.1 and compute spectrophotometry from the reddened spectrum.  This photometry represents what would have been observed.  We then convert the photometry to flux using the standard conversion factors discussed above and then correct for extinction by unreddening the SED points by the same extinction law.  The top panel of Figure \ref{fig_redseds} shows the overcorrection which occurs when the standard flux conversions are used on the reddened SED and then corrected for extinction.  We emphasize that the error is not in the extinction parameters but how they were applied in the analysis.  The bottom panel shows the result of fitting an SED to  the reddened points and then correcting the SED for the extinction.  This gives a much better agreement to the original spectra after the reddening correction.  

Correcting the magnitudes can be done accurately if the spectral shape and extinction curve over the whole filter are considered.  Conversion factors applicable to different sources can be computed (e.g. \citealp{Brown_etal_2010}) but there are still variations and the extinction correction terms are non-linear for significant reddening in the UV. The preferred approach is to redden a spectroscopic model and compare it to the observed photometry (see e.g. the application to the reddened SN~2014J in \citealp{Amanullah_etal_2014, Foley_etal_2014, Brown_etal_2014J}).  Alternatively, one can convert the photometry into flux to create an SED consistent with the photometry, and then deredden the SED rather than computing corrections at individual points.
The uncertainty in the integrated luminosity can be estimated by varying the applied extinction and comparing the output.  The effect may not be linear.  The extinction correction and uncertainty would also vary by epoch, as the changing SED will result in different total extinction even if the amount of dust causing the reddening is constant.

\section{Bolometric Flux Estimates} \label{bolo}

In this section we will describe some of the choices to be made when computing bolometric fluxes or luminosities and then test their effect.  Some of these will be shown in Figure \ref{fig_seds} which compares SEDs made using different methods to the input spectra of Vega and SN~1992A with the area accounted for in the bolometric flux integration shaded.  The differences will be quantified in Section \ref{comparisons}.

\subsection{The Limits of Bolometry--Integrated Flux Measurements} \label{bololimits}

In practice one cannot observed the total bolometric luminosity of an object but are restricted to certain wavelength/energy ranges due to instrumental and atmospheric limitations.  
Integrated flux or luminosity values based on observational data are often given the labels ``pseudo-bolometric' or ``UVOIR'' ( for ultraviolet-optical-infrared) luminosity to indicate that they are not covering the entire energy span.  These terms, however, should be considered adjectival rather than definitive, because they do not clarify the wavelength range actually covered.  Even with the term UVOIR, the UV usually only represents the ground-based near-UV Johnson U or sloan u bands, and the IR usually means some NIR data, which may include the JHK bands or even just the I band.  Others more explicitly report the bands used, such as L$_{UBVRI}$, L$_{BVR}$, etc.  This is a large improvement, because it allows data of the same range to be compared.  We recommend a step further, namely explicitly specifying the wavelength bounds of the integration.  We suggest this be done as L$_{[1600-6000\AA]}$ (see e.g. \citealp{Fransson_etal_2014}).  This might more accurately be called an integrated flux or luminosity.   Integrating between defined wavelength limits and reporting the values by wavelength range rather than just filter could also improve comparisons between objects observed using different systems or at different redshifts.  Using appropriate instrumental calibrations, photometrically consistent SEDs can be constructed and integrated over a common wavelength range.  The uncertainties inherent in the instrumental passbands and photometric calibration will still exist, but there is a reduced uncertainty from converting from one system into another and then performing the bolometric procedure.  \citet{Lyman_etal_2014} discuss the conversion of L$_{ugriz}$ to L$_{UBVRI}$.

It is often desirable to compare the integrated luminosity of objects for which different wavelength ranges were observed.  Assumptions about the missing flux are often made to expand a pseudo-bolometric L$_{BVRI}$ into  L$_{UBVRI}$.  With increased numbers of SNe observed in the UV with Swift/UVOT, comparisons are often desired between recent SNe with UV observations and historical SNe without such observations.  We caution that such extrapolations might eliminate the usefulness of the comparisons.  If there is interesting temporal behavior or a significant amount of UV flux, then questionable assumptions would have to be made to add that flux in (though in some instances the assumptions might be justified).  If the UV flux is low or considered to be well understood, then adding it in does not seem to add anything useful. Comparisons between objects should probably be restricted to the observed wavelength regions in common between the objects.  Similarly for comparisons with models, they should be integrated over the same range.  If there is agreement between the observed luminosities in the filters, then this gives more confidence in the bolometric luminosity.  Theoretical models which do not have a spectral prediction to integrate may require assumptions and extrapolations to be made to either the models or the observations.

Given a set of observations, one must choose between which wavelengths to integrate the flux.  One could use the full range of the filters or more conservatively integrate between the effective wavelengths of the filters at either end.  One could question whether it is appropriate to use the full range of the UV filters (especially those with a red leak) for very red sources or even exclude those filters altogether.  It is true that that for very red objects there is not much UV emission and that it is relatively less constrained due to the number of optical photons contributing to the observed magnitudes of a red source.  The fact that there is little UV emission is information already, and with multiwavelength observations the amount can be constrained. \citet{Ergon_etal_2014} modeled the contribution of the red tails for uvw2 and uvw1 for SN~2011dh.  They wound up excluding the uvw2 and uvw1 filters because of the large optical contamination, though the calculation of the contamination already tells one how much flux comes from the UV. The same principle holds for the tails of the filter transmission at either end.  The tails have less weight, but if there is significant flux in the regions covered by the tails, then those photons would contribute significantly.  Multi-wavelength observations which constrain the SED over the wavelength range of the filter allows one to account for all of the photons regardless of where they come from.  The creation of multiple spectra which are consistent with the observations would allow one to properly estimate how much contribution could come from the filter tails.  
 In this work we use the full range of the UVOT from uvw2 and v, namely 1600-6000 \AA.  To compute spectrophotometry we create spectra and SEDs covering the full range of the tabulated filter curves, 1600-8000 \AA.  Some of the filters have a tiny amount of transmission at those wavelengths.  Our SEDs are extrapolated with a constant flux from 6000 to 8000 \AA, but this has no significant effect on our results due to the small transmission. 

When the flux is integrated between certain bounds (and not being extrapolated), one must clarify what those bounds mean, namely how one deals with the endpoints.  Often it is noted that the flux is set to zero outside the limits of integration.  This in itself does not matter because by definition the flux outside the limits is ignored.  But often the flux endpoints of the integration are set to zero, rather than just the flux outside of the integration.  
This was often done for the UV flux of SNe Ia, for which observations from IUE showed the UV flux was much smaller than the optical \citep{Suntzeff_2003}.  In those cases the effect was small, but the practice has continued for SNe with significant UV flux (e.g. \citealp{Inserra_etal_2013}).  
Since the flux density assigned to a given filter is roughly the average flux density in that filter bandpass for a relatively flat spectrum, setting the boundary point to zero will undercount the flux in that filter by about twenty-five percent.  The total affect this has on the bolometric luminosity depends on the spectral shape and the number of filters being integrated over.  
The top panels of Figure \ref{fig_seds} (labeled Standard-Zero) shows an SED computed from the standard flux conversion factors (stellar average from \citealp{Poole_etal_2008}) with the end points set to zero flux.

A more reasonable assumption would be to set the endpoints to the same flux as the nearest point, essentially assuming a constant flux between the effective wavelength and the integration bound.   
The second panels of Figure \ref{fig_seds} (labeled Standard-Flat) shows an SED computed from the standard flux conversion factors with the end points set to the same value as the nearest filter.  The endpoints will sometimes overestimate and sometimes underestimate the flux, but it does not systematically underestimate the flux by assuming zero flux.  An even better approach would be to set the endpoints to a value such that the SED is photometrically consistent with the photometry.  This is discussed more below.

\subsection{Flux Conversion and Integration for Integrated Flux Measurements} \label{boloflux}

When flux-calibrated spectra covering a large wavelength range are available, a pseudo-bolometric flux can be determined by integration under the spectrum \citep{Hallock_1895, Code_etal_1976, Panagia_etal_1980,Wang_etal_2012,Pereira_etal_2013,Pan_etal_2015}.  
In the absence of flux-calibrated spectra, the flux is estimated based on photometric measurements.  We now return to the concept of flux conversion and examine the effect on integrated flux measurements.
In Section \ref{conversion} we argued that comparisons of photometry with model spectra was most naturally done in units corresponding to the observations (count rates or direct conversion into magnitudes).  For a visual comparison, flux conversion factors can be individually computed (as in \citealp{Brown_etal_2010}) from a model spectrum to look for consistency between the flux from the model spectrum and that derived from the photometry.  However, for the purposes of flux integration those custom conversion factors may not be appropriate because the effective wavelengths may fall on emission or absorption features, leading to an under or overestimate of the flux.

Instead of calculating flux integration points directly from the photometry by assuming conversion factors, one needs an SED consistent with the photometry.  One approach would be to start from the photometry (and converting to flux using any technique as a starting point) and adjust the points to create an SED which is consistent with the measurements.  This is similar to the photometric method described in \citet{Ergon_etal_2014}.  Iteration is necessary because the lines connecting two neighboring filters affect the flux in both of those filters (and possibly others in the presence of filter leaks).  This allows the broad continuum shape to be incorporated into the conversion process.  As such SEDs have to be run through the filter curves, this is a computational, rather than purely analytic, process.  In one such algorithm, flux points at the Vega effective wavelengths can be iteratively adjusted one at a time to be consistent with the multi-band photometry.  For the UVOT photometry, we find it most effective to begin at the optical end of the spectrum.  Once that region is approximately known, the red tails of the UV filters are appropriately accounted for and the UV flux can be determined.  The flux at the end points can also be varied to minimize the photometric differences between the observations and the assumed SED.
Such SED reconstructions are shown in the fourth panels of Figure \ref{fig_seds}.  This SED is constructed using points at each of the effective wavelengths of the four interior filters and the endpoints at 1600 and 6000 \AA.  Since the flux level of the points is determined by an iterative comparison with all of the photometry, it is not necessary for there to be fixed wavelengths or a one-to-one relationship between the SED points and the wavelengths of the filters.  More complicated methods could add as many wavelength points as necessary to match the complexity required by the photometry.  The computational time for a grid search of SEDs would scale as $f^w$, where f is the number of flux points and w the number of wavelength points.  
SEDs made from somewhat arbitrary line segments seem less strange when noting that flux conversions for a wide variety of objects use historical factors based on stellar templates very unlike the objects under question anyway.  

Continuing in complexity, a spectral template that may be similar to the object in question may be used, with wavelength dependent scaling or color-matching (sometimes referred to as wavelength-dependent warping) to bring the spectrum into agreement with the observed broad-band photometry \citep{Howell_etal_2009}.  In the bottom panel of Figure \ref{fig_seds} we choose from our full set of spectra the fifth best match in uvm2-uvw1 and uvw1-v colors (so that we do not just pick the identical spectrum) and modify it to best match all of the photometry.  This modification, sometimes referred to as ``warping,'' ``mangling'' or ``color-matching'' is done by finding a best-fit SED using the grid search above and linearly interpolating a scale factor between the two SED fits.  This scaling is applied to the template spectrum and iterated as needed.  Fitting the spectra in such a way accounts for the optical contribution to the UV filters.  Creating a scaling function from the count rates themselves and the effective wavelengths ignores the optical contribution to the UV filters and makes it hard to match the observed count rates without many iterations which can drive portions of the spectra to arbitrarily large or small values in an attempt to fit.   While we apply a linearly-interpolated scaling to match the spectral template to the input count rates, one could use low order polynomials, splines, or physically-motivated functions such as a reddening law (e.g. \citealp{Cardelli_etal_1989} used in \citealp{Nugent_etal_2002} for optical data), Lyman-alpha breaks at various redshifts, or metallicity-dependent flux ratios \citep{Foley_Kirshner_2013}.  There will of course be degeneracies as different spectral shapes, features, and color-matching functions could result in the same observed magnitudes.  A solution may not be unique, however, but at least the SED would be consistent with the photometry.  Utilizing a large set of differing SEDs which are nonetheless photometrically consistent might be a way to gauge the accuracy of an integrated flux measurement. 
 The first goal is to create a spectrum consistent with the observed photometry before expecting anything made from the spectrum to be accurate. The ideal case is a spectrophotometrically accurate spectrum covering a large wavelength range (see \citealp{Wang_etal_2009_05cf} and \citealp{Pereira_etal_2013} for well observed SNe approaching this ideal).  Spectra with a smaller wavelength range could also be incorporated \citep{Ergon_etal_2014} utilizing photometry or spectral templates at the epochs/wavelengths not covered spectroscopically.
Once the SED or spectrum is consistent with the photometry we can start to believe that the flux integrated under that curve might be accurate.

\section{Quantitative comparisons} \label{comparisons}

\begin{figure*} 
\resizebox{14cm}{!}{\plottwo{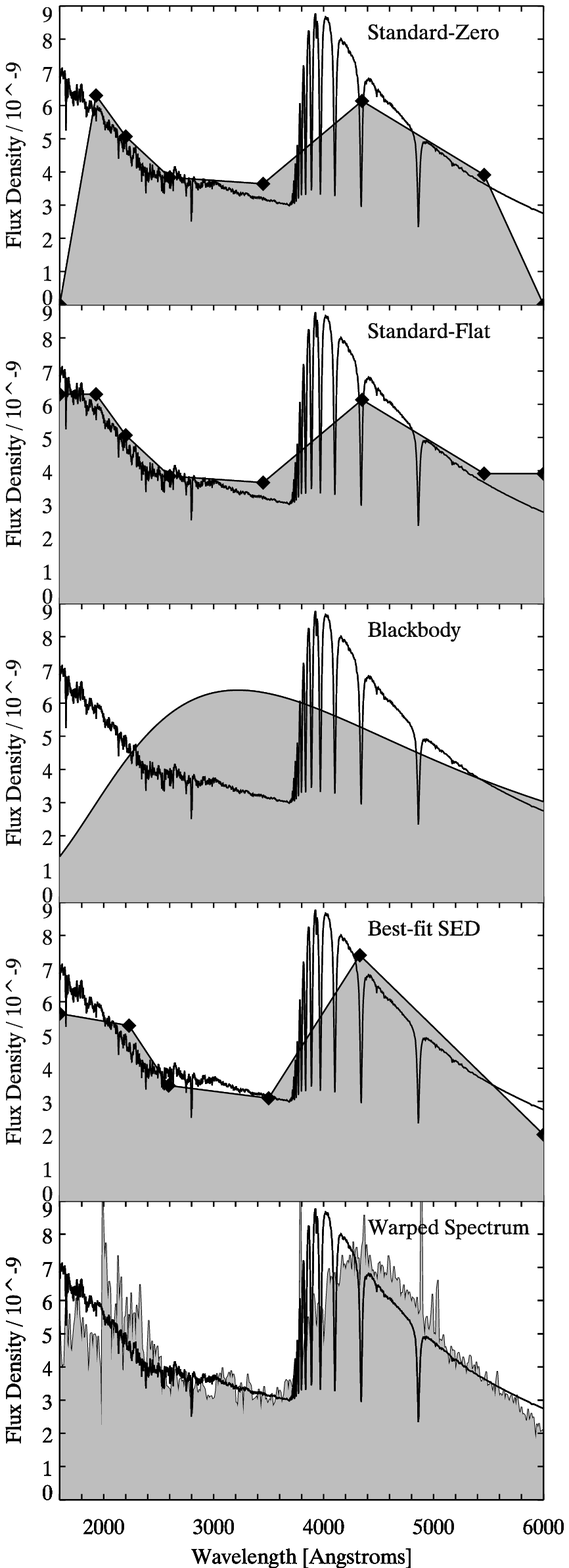}{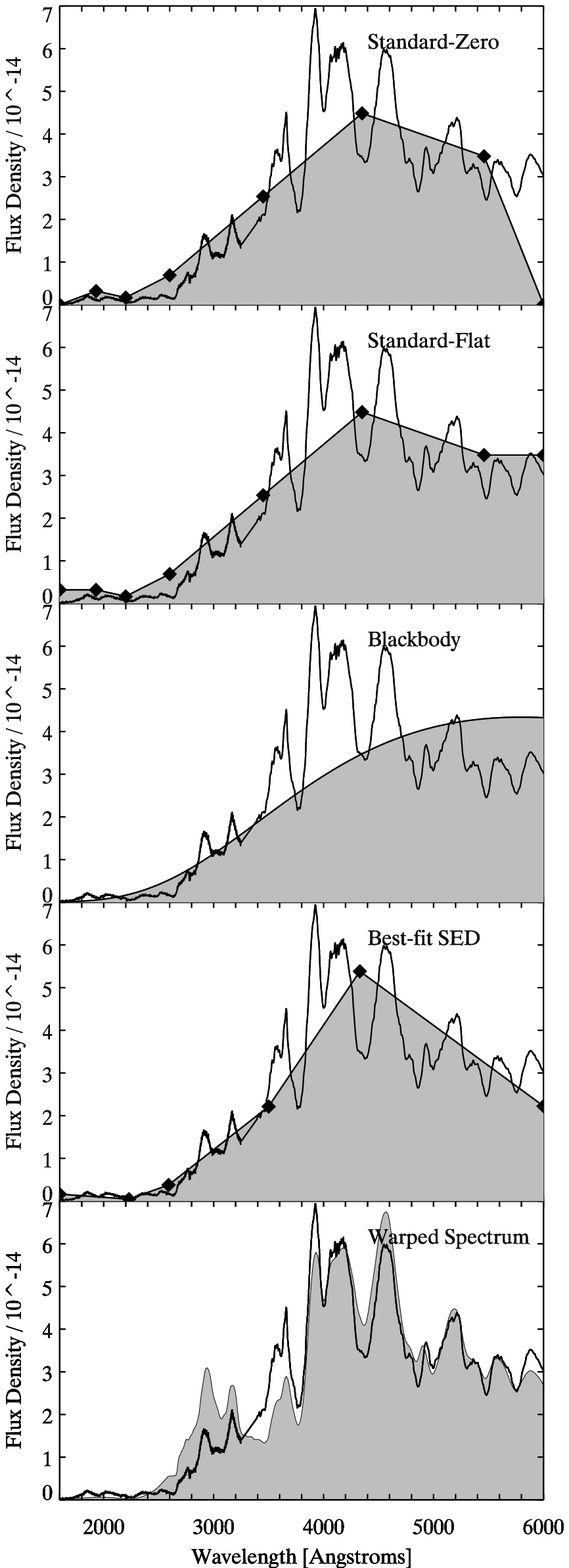}  }
\caption[Results]
        {SED reconstructions (in units of ergs s$^{-1}$ cm$^{-2}$ \AA$^{-1}$)are shown for Vega (left panel) and SN~1992A (right panel).
The different rows show SEDs with diamonds reconstructed using the methods described in the text, with the corresponding integration of the flux shown as the shaded region.

 } \label{fig_seds}    
\end{figure*}

\begin{figure*} 
\resizebox{16cm}{!}{\includegraphics*{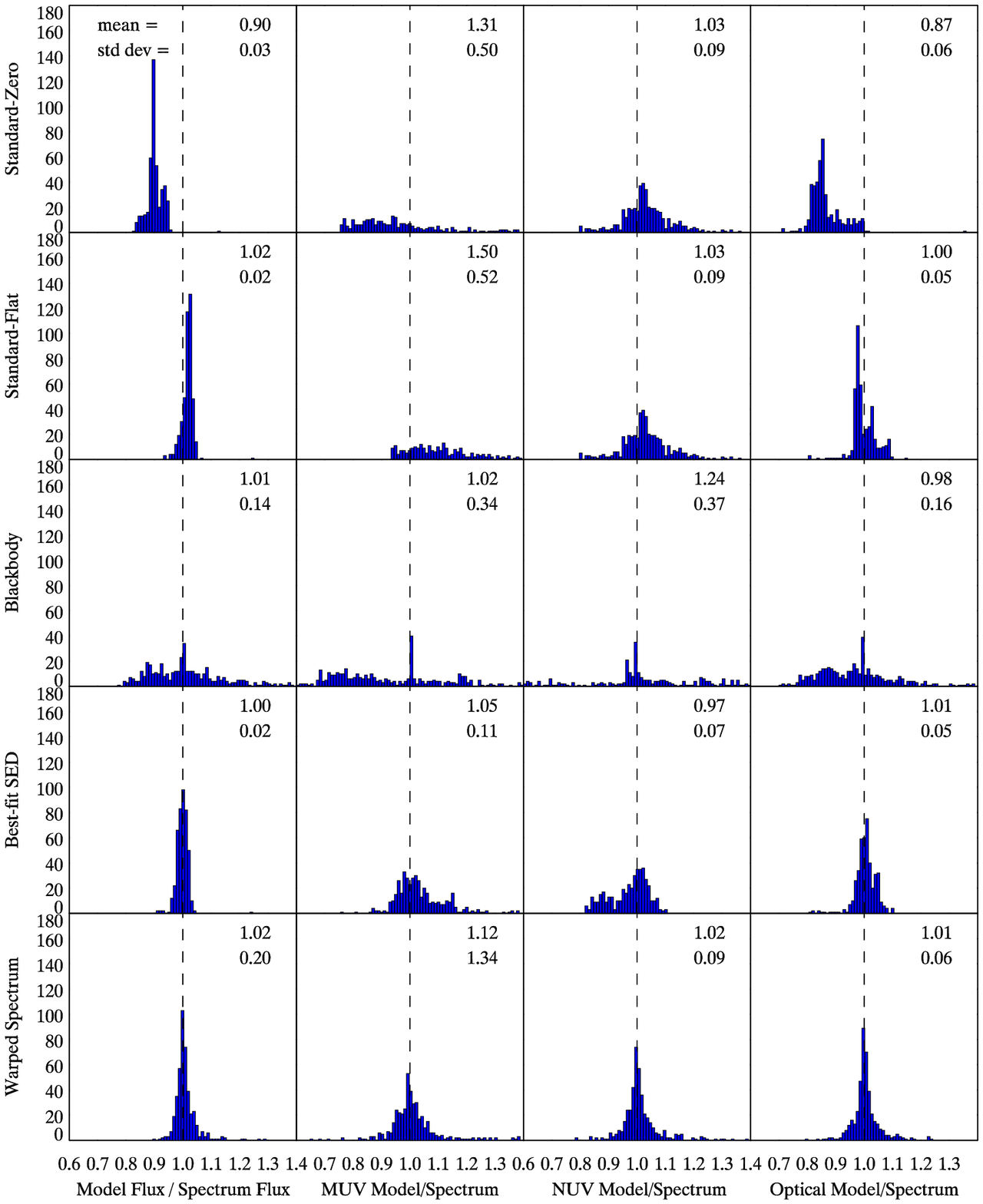}   }
\caption[Results]
        {Histograms of the ratio of the integrated luminosity from the SEDs to the ``true'' integrated luminosity from the original spectrum from different wavelength regions.  Left: mid-UV (1600-2800~\AA), middle: near-UV (2800-4000~\AA), right: optical (4000-6000~\AA).  The rows correspond to the SEDs described in the text and displayed in Figure \ref{fig_seds}.  Shown for each are the mean and standard deviation, though these can be highly affected by a few outliers.
 } \label{fig_bolohist}    
\end{figure*}

While one does not know a priori what the intrinsic spectrum of an observed source is (or one would not be trying to estimate the bolometric flux from the photometry), one can test how well different methods work for a large variety of test spectra.  
To quantitatively test the effect of these assumptions we use a large sample of input spectra, including stellar spectra from \citet{Pickles_1998}, galaxy spectra from \citet{Storchi_etal_1995}, a type Ia spectral template series from \citet{Hsiao_etal_2007}, theoretical spectra matched to the SN~IIP 2005cs \citep{Dessart_etal_2008}, and blackbody spectra with temperatures ranging from 2,200-38,000 K. 
We measure the differences between the modeled SED and the original spectrum by integrating the flux between 1600 and 6000 \AA~(the effective UVOT range) and in subregions from 1600-2800 \AA~(mid-UV or MUV), 2800-4000~\AA~(near-UV or NUV), and 4000-6000~\AA~(optical, in this case covering the B and V bands).  The ratios of the SED flux to the original spectrum flux are displayed as histograms in Figure \ref{fig_bolohist}.  The columns correspond to the integrated, MUV, NUV, and the optical flux ratios (model divided by actual).   The histogram rows correspond to the same models shown in Figure \ref{fig_seds}.  Next we will review those models and comment on the results.

$\bullet$ Standard-Zero --  SED is computed from the standard flux conversion factors (stellar average from \citealp{Poole_etal_2008}) at the filter effective wavelengths with the end points set to zero flux.  The effect of setting the end points to zero results in a systematic underestimate of the flux in the MUV and optical portions (the regions covering the ends). 

$\bullet$ Standard-Flat --  SED is computed from the standard flux conversion factors (as above) with the end points set to the same flux as the neighboring filter.  The systematic underestimate is removed.

$\bullet$  Blackbody -- A grid of blackbody spectra is searched for the temperature at which a Planck spectrum gives the smallest difference between the observed and predicted count rates.  

We do not consider this further, as the intrinsic spectrum is not generally known in real-life application; if a spectrum is available, one should integrate beneath it to obtain the flux rather than a photometric SED.  We also point out that having an SED with the exact flux densities corresponding to given wavelengths as the corresponding spectrum is not appropriate because the linear interpolation between the points may still give the incorrect integrated flux especially if there are strong absorption or emission features at the chosen wavelengths.

$\bullet$ Best-fit SED --  This SED is constructed using wavelength points at each of the effective wavelengths of the four interior filters and the endpoints at 1600 and 6000 \AA.  A grid of flux values at those points is created, tested, and modified to minimize the difference between the input and computed six-filter count rates.  This is done without any prior knowledge of the spectral shape.  Forcing the SED to agree photometrically results in a much better agreement with the UV flux values.  

$\bullet$ Warped Spectrum --  A spectrum with similar (but not exact) colors is chosen from the test spectra and adjusted to minimize the difference between the six filter input magnitudes and the computed spectrophotometry.   The significant improvement in the UV is an indication that the complex spectral shape in the UV is poorly fit by crude SEDs.

Figure \ref{fig_bolohist} shows the mean and standard deviation of the flux ratios, though these values should be used cautiously if at all 
as these are dependent more on the sample spectra tested than the methods.  For example, the use of cool stellar SEDs or cool blackbodies can result in no MUV flux, driving certain parameters to zero or infinity and making it difficult to derive sample properties without arbitrary cuts in color or standard deviations for the mean.  For particular science questions, one can estimate the systematic shift and/or spread in measurements by using a reasonable set of simulated spectra applicable to the measurement being made.

These simulations lead to three comments.
First, there is no reason to arbitrarily set the flux at any boundary to be zero.  If the observations require negligible or zero flux then that would be correct.  Second, the failure of the standard flux conversions in the UV for very red objects results in a small difference to a broad integrated flux measurement because there is so little UV flux to begin with.  The biggest problems arise if someone is interested primarily in the UV flux, for example to estimate the amount of ionizing flux incident on circumstellar gas and dust \citep{Simon_etal_2009}.  The third conclusion is that the standard flux conversion factors can be improved upon with no spectral knowledge.  From the broad-band photometry alone one can iteratively reconstruct the SED to get a more accurate understanding of the spectral energy distribution.  This can be further improved by utilizing spectral templates which are at least similar to the source in question.  Matching the smaller scale features in the spectrum can help improve our broad-band understanding of the flux from SNe (and other objects).


\section{Integrated flux curves of supernovae} \label{snbolo}

\begin{figure*} 
\plottwo{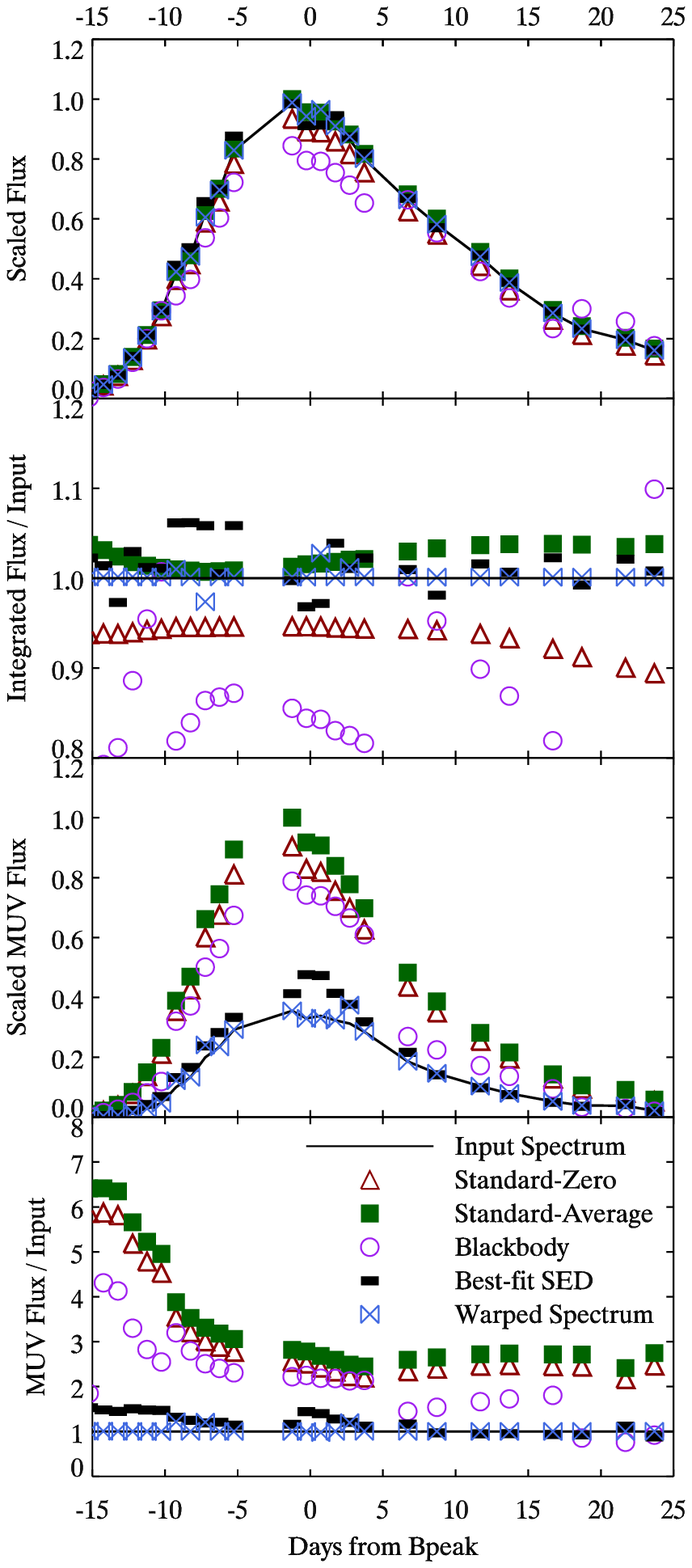}{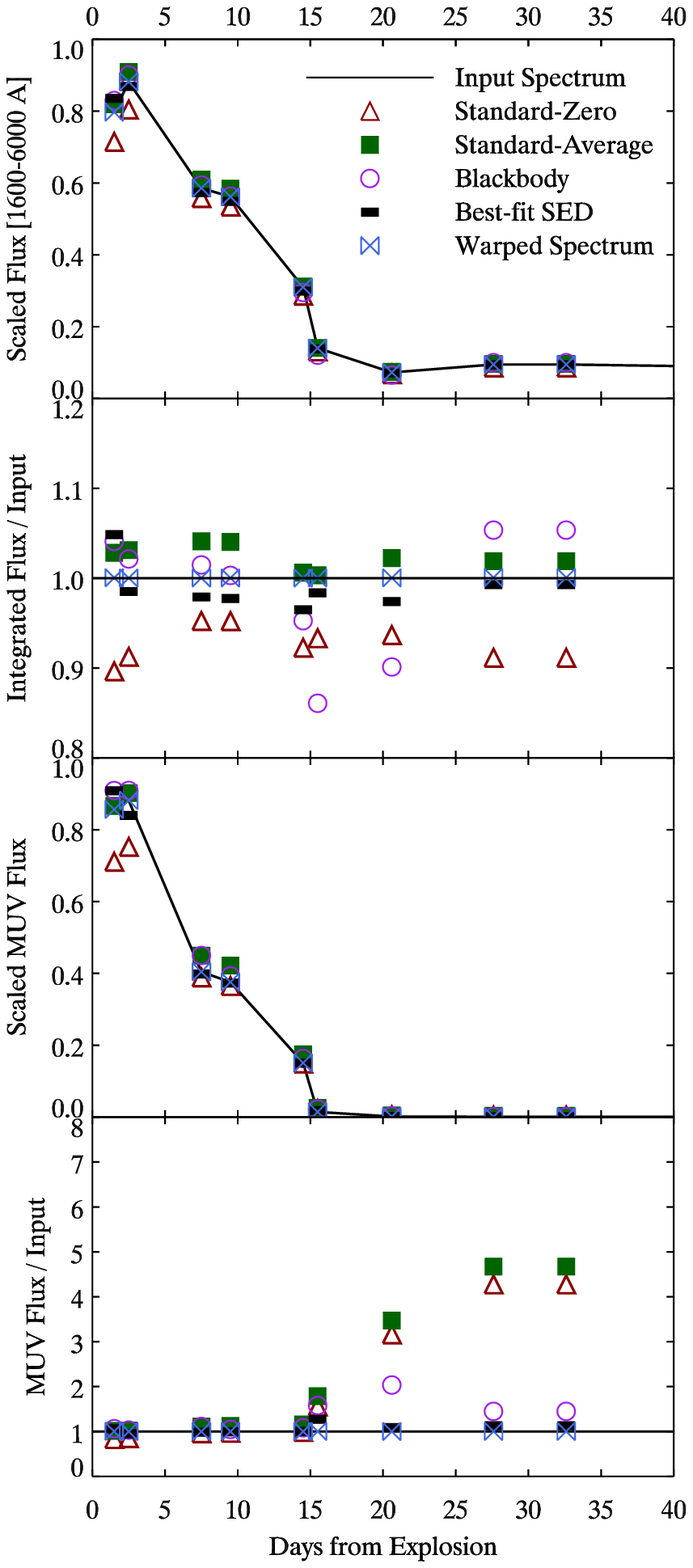}  
\caption[Results]
        {Top Left: Comparison of bolometric lightcurves using different methods for the bolometric spectral series of SN~2011fe \citep{Pereira_etal_2013}.
Second Panel: Ratio of the derived integrated flux from different methods to the flux integrated directly from the spectrum.
Third Panel: Integrated mid-UV flux using different methods.
Fourth Panel: Ratio of the derived integrated mid-UV flux compared to the mid-UV flux integrated directly from the spectrum.
Right panels: same as the left for theoretical model spectra of SN~2006bp \citep{Dessart_etal_2008}.
 } \label{fig_snbolos}    
\end{figure*}

Having tested the effects of different methods on an arbitrarily large set of spectra, we now explore the effects for the integrated flux of SN models.  We use two spectral series as the `truth table' against which to compare the outputs of the different methods.  For a type Ia SN we use the spectral series of SN~2011fe from \citet{Pereira_etal_2013}.  At each epoch we measure the integrated flux in the 1600-6000 \AA~region as well as subregions.  We also compute spectrophotometry in the UVOT system.  From this synthetic photometry, we convert the magnitudes into flux densities and integrate the flux over the same regions.  For a similar spectral template we use the HST UV/optical spectrum of SN~1992A which was the standard for many years \citep{Kirshner_etal_1993}.  In the top left panel of Figure \ref{fig_snbolos} we show the integrated flux from the original spectral series and each of the SED reconstructions.  The second panel down shows the ratio of the calculated flux to the actual flux.  
While the Standard-Zero SED systematically underestimates the flux and the Custom-Flat SED over or underestimates the flux by 10\%, the others all match the flux to within 5\%.  The third panel down shows the mid-UV flux (1600-2800 \AA) and the bottom panel the ratio of the calculated flux to the actual mid-UV flux.  All of the models overestimate the flux at early times.  Those using the standard flux conversion factors are a factor of $\sim6$ too high due to the optical contamination of the uvw2 count rates.  The warped spectrum does the best job, as starting with a similar spectrum to get the overall spectral shape correct does a better job than five linear segments even when both can be made to match the observed magnitudes.

The right panels of Figure \ref{fig_snbolos} show the same thing for a theoretical spectral series matched to the type IIP SN~2006bp \citep{Dessart_etal_2008}.  SN~2006bp is quite different because it has a strong color evolution.  The effect is greatest in predicting the mid-UV flux.  At early times when the mid-UV flux is high the models are close to the correct flux.  As the SN reddens, however, the optical contamination causes the standard flux conversions to fail dramatically in the UV.  A warped UV-optical spectrum of SN~1999em from a few weeks after explosion \citep{Baron_etal_2000} is able to better match the late mid-UV flux than a linear SED.

\begin{figure*} 
\resizebox{18cm}{!}{\includegraphics*{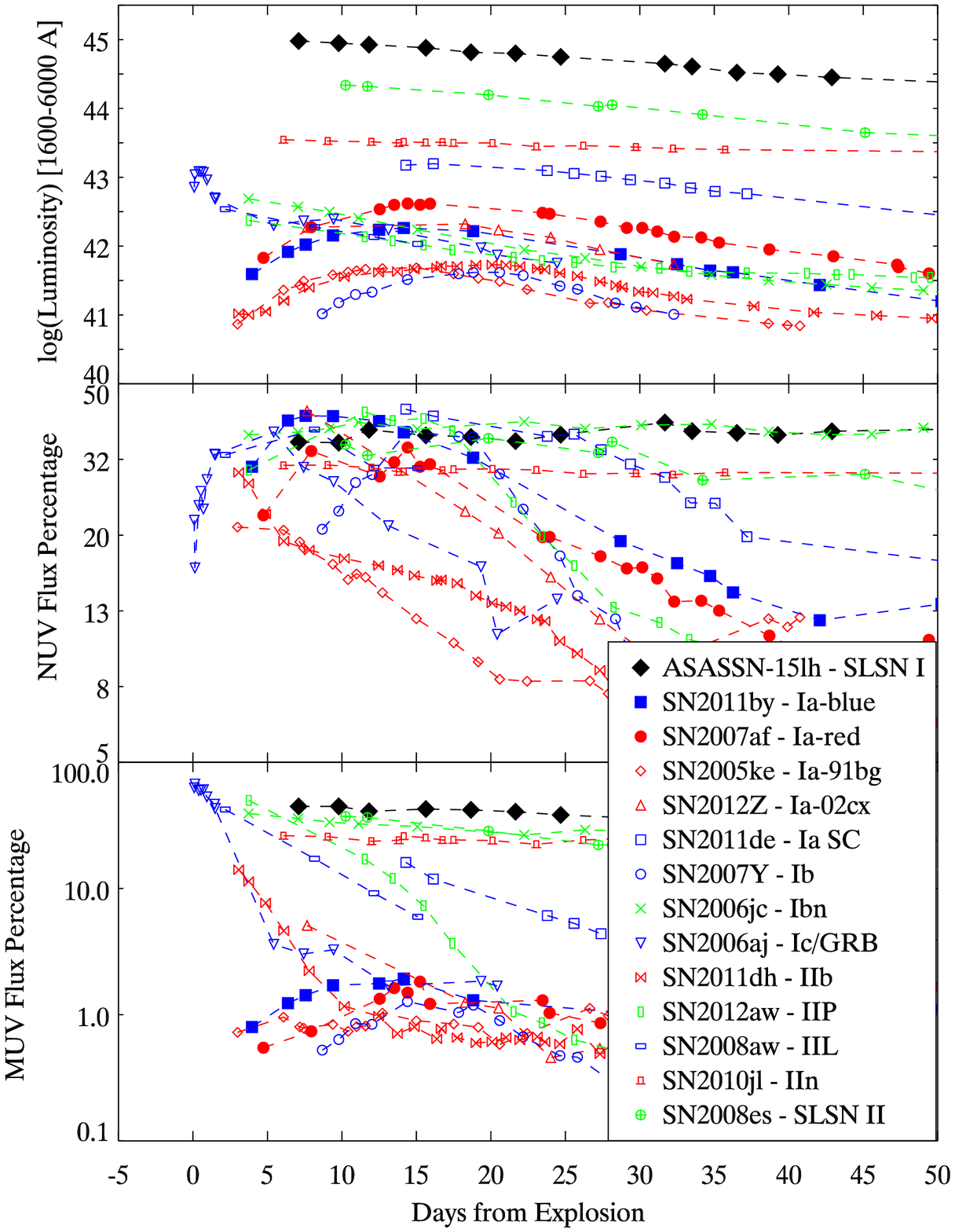}   }
\caption[Results]
        {Top Panel:  Integrated luminosity curves of representative SNe of different types given in units of ergs s$^{-1}$ (over the UVOT wavelength range of 1600-6000 \AA).  Lower panels: the fraction of the luminosity coming from the NUV  (2800-4000 \AA) and MUV (1600-2800 \AA).
 } \label{fig_allsnbolos}    
\end{figure*}

Finally, in Figure \ref{fig_allsnbolos} we use the Best-fit SED method to compute integrated flux measurements for SNe representing most SN classes and subtypes (see \citealp{Brown_etal_2014_SOUSA} for more details).  Also plotted are the fraction of the 1600-6000 \AA~flux coming from the mid-UV (1600-2800 \AA) and near-UV(2800-4000 \AA) wavelength regions.  SNe vary greatly in their luminosities and colors.  To properly understand SNe, we need to understand the observations themselves.  These UV-optical SEDs can then be applied to understand the rest-frame UV-optical properties of high redshift SNe.

\section{Conclusion}\label{conclusion} 

In summary, we recommend the following principles for understanding the flux from photometrically observed sources:

$\bullet$ Model spectra are best compared with photometric observations by  forward modeling the spectrum with assumed reddening and appropriate photometric calibration to compare with the observed count rates or magnitudes.

$\bullet$ Interpreting heterochromatic broad-band measurements as monochromatic flux densities must be done with great care and understanding of the photometric systems and the intrinsic spectral shape.

$\bullet$ Extra care must be taken to match the SED of a reddened spectrum before correcting it for extinction, as small errors in the assumed UV SED translate into large errors after extinction correction.

For the integration of flux or luminosity we recommend the following.

$\bullet$ An SED should be made which is consistent with all available observations rather than just computing individual flux density points and connecting the dots.  Flux-calibrated spectrophotometry would be ideal.  Color-matching a similar spectral template to photometry is the next best choice.  Reconstructing a simple SED from a few straight line segments to be consistent with the photometry, however, is already an improvement from constructing straight-line SEDs from average flux conversion factors.

$\bullet$ The limits of integration should be explicitly defined, e.g. L$_{\mathrm{1600-6000\AA}}$, and the flux density at the endpoints should be based on a photometrically-accurate SED or spectrum rather than arbitrarily set to zero.

$\bullet$ Comparisons of integrated flux between objects should be restricted to the observed wavelengths common to all objects.

\vspace{10mm}

\section{Acknowledgements}\label{acknowledgements} 

We appreciate helpful comments on the manuscript from M. Stritzinger.  The development of this manuscript benefitted greatly from discussions with N. Suntzeff, the Texas A\&M Aggienova Group, and the Swift/UVOT team.  This study was performed and published to encourage proper use of data from the Swift Optical/Ultraviolet Supernova Archive (SOUSA).  SOUSA is supported by NASA's Astrophysics Data Analysis Program through grant NNX13AF35G.

\clearpage
%
%
%
%
%
%
%
%
%
%
%
%
%

\bibliographystyle{apj}

\end{document}